%% file: srds_2021.tex
\documentclass[10pt,conference,a4paper]{IEEEtran}
\input{header}

\begin{document}

\title{Dandelion: multiplexing Byzantine agreements to unlock blockchain performance}

\author{
 \IEEEauthorblockN{
     Kadir Korkmaz\IEEEauthorrefmark{1}
     Joachim Bruneau-Queyreix\IEEEauthorrefmark{1}
     Sonia Ben Mokthar\IEEEauthorrefmark{2}
     Laurent Réveillère\IEEEauthorrefmark{1}
}

 \IEEEauthorblockA{\IEEEauthorrefmark{1} Univ. Bordeaux, Bordeaux INP, CNRS. LaBRI, UMR 5800, Talence, France }%
 \IEEEauthorblockA{\IEEEauthorrefmark{2} CNRS - LIRIS. Lyon, France}%

}
\maketitle


\input{abstract}


\input{sections/intro}

\input{sections/model_and_goal}
\input{sections/algorand}

\input{sections/overview}

\input{sections/implementation}

\input{sections/evaluation.tex}

\input{sections/sota}

\input{sections/conclusion}

\balance
\bibliographystyle{IEEEtran}
\bibliography{biblio,bbt}

\end{document}

%% file: header.tex
\usepackage[T1]{fontenc} 
\usepackage[utf8]{inputenc}

\usepackage{cite}
\usepackage{amsmath,amssymb,amsfonts}
\usepackage{algorithmic}
\usepackage{graphicx}
\usepackage{textcomp}
\usepackage{xcolor}

\usepackage{xspace}
\usepackage{algorithmic}
\usepackage{tabularx,booktabs}
\usepackage[ruled,vlined]{algorithm2e}

\usepackage{amsthm}
\usepackage{siunitx} 

\usepackage[ddmmyyyy]{datetime}
\usepackage{lastpage}
\usepackage{fancyhdr}
\usepackage{balance}
\usepackage[abbreviations]{foreign}
\usepackage{pifont}

\newcommand{\isep}{\mathrel{{.}\,{.}}\nobreak}

\newcommand{\protoname}{\textsc{Dandelion}\xspace}
\newcommand{\ba}{\textit{BA$\star$}\xspace}
\newcommand{\bba}{\textit{Binary}\ba}
\newcommand{\dba}{\textit{Dandelion BA$\star$}\xspace}
\newcommand{\dbba}{\textit{Dandelion Binary}\ba}

\newcommand\dingnum[1]{\ding{\numexpr181+#1}}

%% file: abstract.tex
\begin{abstract}
    Permissionless blockchain protocols are known to consume an outrageous amount of computing power and suffer from a trade-off between latency and confidence in transaction confirmation. 
    The recently proposed Algorand blockchain protocol employs Byzantine agreements and has shown transaction confirmation latency on the order of seconds.
    Its strong resilience to Denial-of-Service and Sybil attacks and its low computing power footprint make it a strong candidate for the venue of decentralized economies and business ecosystems across industries.
    Nevertheless, Algorand's throughput is still far from the requirements of such applications. 
    In this paper, we empower Algorand's protocol by multiplexing its byzantine agreements in order to improve performance.
    Experiments on wide area networks with up to ten thousand nodes show a 4-fold throughput increase compared to the original Algorand protocol.
\end{abstract}

%% file: sections/intro.tex
\section{Introduction}\label{sec:introduction}

Blockchains are decentralized, globally distributed, strongly consistent replicated systems that run across networks of mutually untrusting nodes. 
Since the emergence of the decentralized Bitcoin protocol~\cite{nakamoto_bitcoin_2008}, permissionless blockchains have demonstrated their ability to run arbitrary distributed applications~\cite{androulaki_hyperledger_2018}, with the promise of supporting entire decentralized economies~\cite{giladAlgorandScalingByzantine2017} and business ecosystems across industries~\cite{noauthor_hyperledger_nodate}.

Because of the colossal amounts of money involved in these systems of growing interest, and not just from honest parties, permissionless blockchains are more and more at the center of attacks~\cite{lin2017survey}.  
Attacks by Sybil participants (\ie, physical individuals who create a large number of virtual identities) are among the most difficult to identify and counter.
Sybil participants collude to have more impact on collaborative tasks such as voting or to increase their chances of being selected in tasks involving randomness such as random elections~\cite{zhang2019double}.

Therefore, dealing with Sybil attacks in permissionless blockchains has been the subject of active research in the past years. 
A key assumption exploited by existing solutions is that physical individuals rarely have access to unlimited resources. 
Two main resources have been considered in this context: computing resources with proof-of-work protocols (PoW) and financial resources with proof-of-stake protocols (PoS). 
While PoW protocols are unanimously considered as being too energy-intensive, their PoS alternative is becoming increasingly attractive. 
Indeed, besides their lower energy consumption, they also exhibit better performance than PoW-based protocols~\cite{saleh2021blockchain}.

In this context, Algorand~\cite{giladAlgorandScalingByzantine2017} is among the most scalable PoS-based permissionless blockchain.
To scale the consensus to many users, Algorand relies on a cryptographic sortition mechanism that randomly, locally and in non-interactive way selects nodes that are charged to propose blocks. 
It then uses a byzantine agreement protocol called \ba to reach a consensus on one of the proposed blocks.
Algorand has a low transaction confirmation latency of the order of a minute. 
Its evaluation in an emulated wide area network exhibits a 125-fold throughput improvement compared to the Bitcoin protocol.
We show in this paper that despite the impressive performance increase brought by Algorand, this protocol still suffers from performance limitations. 
In particular, its performance drops dramatically with large block sizes.

In Algorand, the time of gossiping blocks in the network always largely dominates the duration of \ba. 
Indeed, disseminating a block as large as 10~MB roughly takes 45 seconds, whereas \ba requires only 15 seconds to complete.
This long block proposal phase is an important limitation to increase the throughput of Algorand transactions. 
Increasing the block size would only increase the confirmation latency and keep the throughput at its highest level in the best case.
To better understand this limitation, we replicated the experimental setup reported by the authors of Algorand~\cite{giladAlgorandScalingByzantine2017} with one thousand nodes and block sizes ranging from 1~MB to 20~MB.
Our implementation of the Algorand protocol is written in Golang and amounts to 3K lines of code. 

\begin{figure}[ht]
    \centering
    \includegraphics[width=\columnwidth]{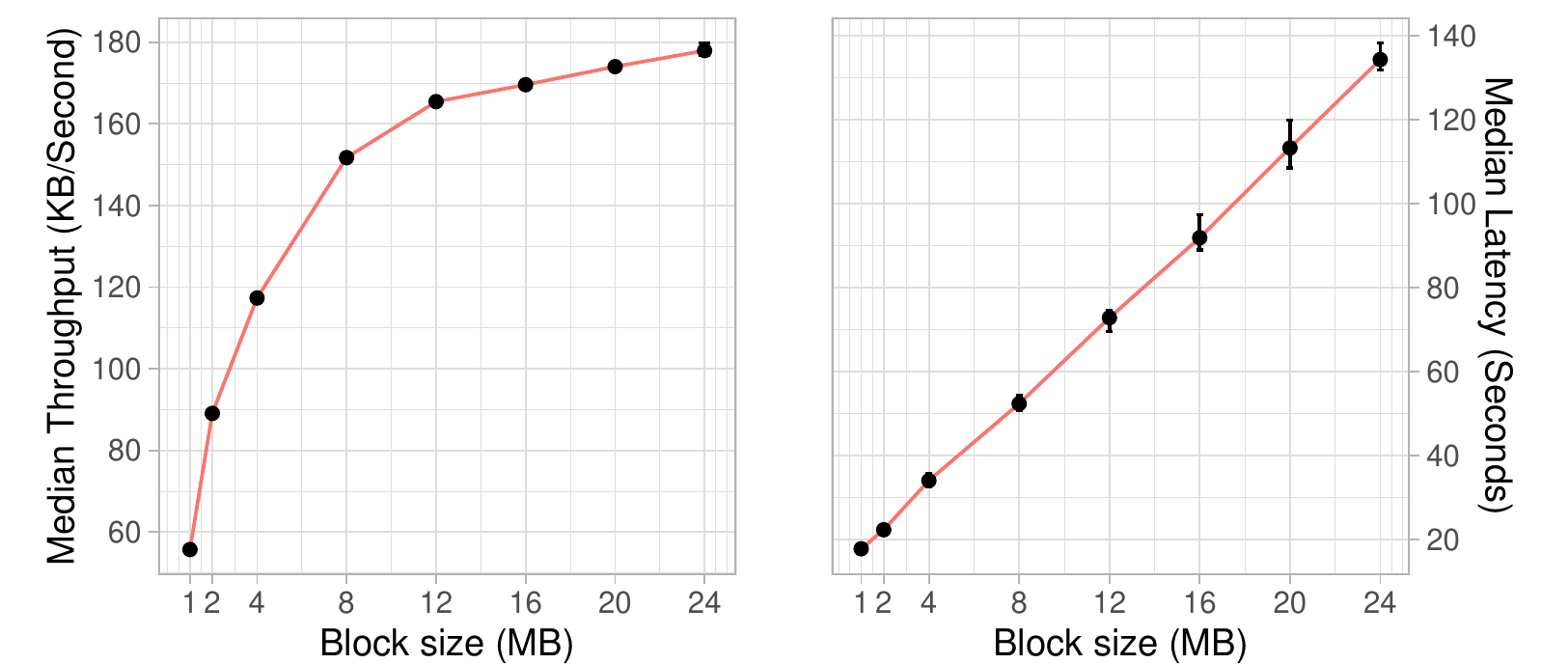}
    \caption{Effective throughput and round latency of Algorand}
    \label{fig:algorand-performance}
\end{figure}

Results reported in Figure~\ref{fig:algorand-performance} show that the latency for appending a block grows linearly with the size of the block reaching almost 2 minutes for 20~MB blocks. 
On the other hand, we observe that Algorand almost reaches its highest throughput with 12~MB blocks. 
Then, very limited throughput gain can be observed with larger blocks.
We noticed in these experiments that although a few dozens blocks are gossiped in the network to extend the chain in each round, only one block is appended, which let us infer a largely underutilized network. 

\bigskip
\noindent
\textbf{Contributions.}
In this paper, we introduce  \protoname, a solution that empowers Algorand by multiplexing its byzantine agreement to improve performance. 
In \protoname, multiple nodes independently propose concurrent blocks containing disjoint sets of transactions. 
They grow the chain by appending a subset of the proposed blocks in every round. 
Since \protoname is built on top of Algorand, it inherits its safety and liveness properties.
Our evaluation involving up to 10,000 nodes deployed on 100 physical machines, shows that \protoname can reach a 4-fold improvement of Algorand’s performance, both in terms of increased throughput and reduced latency.

\smallskip
\noindent
\textbf{Outline.}
The paper is structured as follows. 
We present our system model and assumptions in Section~\ref{sec:model}.
We give a high-level overview of the Algorand blockchain protocol and its building blocks in section~\ref{sec:algorand} before detailing \protoname in section~\ref{sec:approach}. 
We describe our implementation in Section\ref{sec:experimentation} and highlight the settings of our field experiment and the results of our validation in Section~\ref{sec:evaluation}. 
Finally, we discuss related work in Section~\ref{sec:rw} and conclude in Section~\ref{sec:conclusion}.

%% file: sections/model_and_goal.tex
\section{System model and assumptions}\label{sec:model}

\protoname is a Proof-of-Stake committee-based consensus protocol for permissionless blockchain where nodes decide on an ordered log of transaction blocks.
Because we build \protoname on top of Algorand, we inherit the same safety and liveness properties and rely on identical assumptions regarding the network communications and the adversarial model.
More specifically, \protoname provides the following properties:

\noindent
\textbf{Safety}: 
With overwhelming probability, all nodes agree on the same transactions. 
More precisely, if one honest node accepts a transaction, any future transactions accepted by other honest nodes will appear in a log that already contains the former transaction. 
This property holds even for isolated nodes that are disconnected from the network.

\noindent
\textbf{Liveness}:
In addition to safety, \protoname also makes progress under additional network synchrony assumptions that we describe below. 
To achieve liveness, \protoname uses \textit{strong synchrony} where messages sent by
a majority of honest nodes will be received by most other honest nodes in a known time bound. 
This assumption allows an adversary to control a small part of the network but does not allow the adversary to manipulate the network at a large scale and does not allow network partitions.
\protoname also relies on a \textit{weak synchrony} assumption for safety where the network can be asynchronous for a potentially long bounded period if and only if the network recovers in a strongly synchronous state for a reasonably long period again. 
Finally, \protoname assumes loosely synchronized clocks across all nodes, for instance using the NTP protocol~\cite{mills1995improved}, to recover liveness after weak synchrony. 
Specifically, the clocks must be close enough for most honest nodes to initiate a recovery protocol at approximately the same time. 

\protoname relies on standard cryptographic primitives, including public-key signatures and hash functions. 
We assume that money is distributed among nodes. The fraction of money circulating amongst honest nodes is above some threshold $h$ (greater than $2/3$); however, an adversary can participate and own some money. 
We further assume that an adversary can corrupt targeted nodes, but not a large number of nodes that hold a significant fraction of the money.

%% file: sections/algorand.tex
\section{Algorand}\label{sec:algorand}

Algorand requires each node to hold a public and private key pair $(pk_i,sk_i)$. 
Nodes maintain a log of transactions, called a blockchain. 
Each transaction is a payment signed by one node's public key, transferring money to another node's public key. 
Algorand builds the blockchain in successive asynchronous rounds ending with a consensus decision on a possibly empty block.
Each new block contains a set of transactions and the hash of the previous block.

\begin{figure}[hb]
    \centering
    \includegraphics[width=\columnwidth]{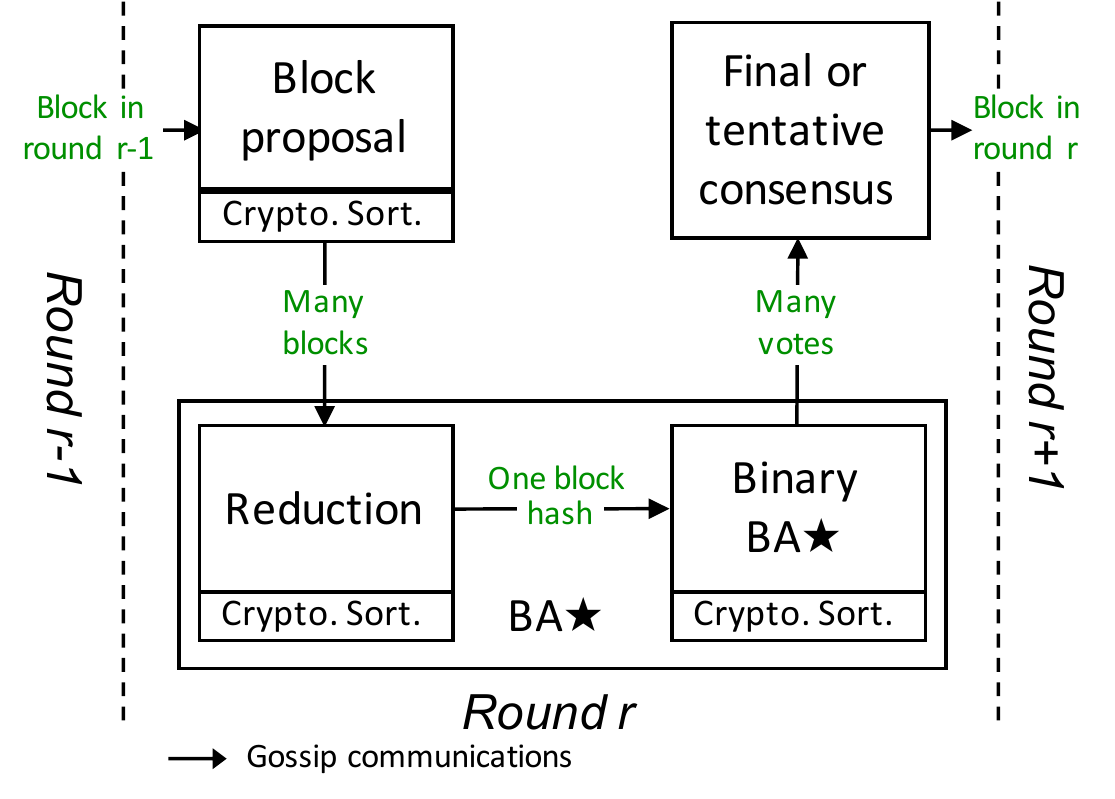}
    \caption{Structure of a round in Algorand}
    \label{fig:algo_round}
\end{figure}

Figure~\ref{fig:algo_round} depicts the four main phases of an Algorand round.
In the first phase, a subset of nodes propose blocks to be appended to the blockchain. 
Then, a byzantine agreement procedure called \ba reduces the problem of agreeing on one of many different block hashes to agreeing on either a selected block hash or an default empty block hash, and reaches consensus via a binary agreement called \bba. 
Every node counts votes cast during the \ba phase to learn about the outcome of the agreement procedure either on a final or a tentative consensus.
On the first three phases of the protocol, committees of nodes are created at random to execute some specific tasks.
To prevent an adversary from learning the identity of committee nodes and forging targeted attacks, nodes independently determine if they belong to specific committees through a cryptographic sortition procedure.
In addition, committees are renewed for each step of the protocol to prevent targeted attacks on committee members once they send a message.

\subsection{Gossip protocol}
Algorand relies on gossip-based communications, with each node selecting a small random set of peers to which it transmits block and protocol messages. 
To ensure that nodes cannot forge messages, messages are signed by the sender’s private key, and each node checks the validity of the signature before relaying it. 
To avoid forwarding loops, nodes do not relay the same message twice.

\subsection{Cryptographic sortition}
Cryptographic sortition enables a small fraction of nodes to be randomly selected for a specific phase of the protocol in a private, local and non-interactive way and weighed by nodes' account balance.
Notably, it provides each selected node with a hash value, which can be compared between nodes, and a proof of the chosen node's selection for a given phase and a priority.

This sortition relies on Verifiable Random Functions (VRF) ~\cite{micaliVerifiableRandomFunctions1999}, a cryptographic primitive that acts as a pseudo-random number generator and that provides publicly verifiable proof of its output correctness.
Informally, on any input string $x , VRF_{sk}(x)$ returns two values: a hash and a proof $\pi$. 
The hash value is uniquely determined by $sk$ and $x$, but is indistinguishable from random to anyone that does not know $sk$. 
The proof $\pi$ enables anyone that knows $pk$ to check that the hash indeed corresponds to $x$, without having to know $sk$.

Each node executes the sortition by computing $\langle hash,\pi \rangle \leftarrow VRF_{sk}(seed || role)$, the VRF's output on a concatenation of a specific role (\eg proposing a block, contributing to a step in the reduction phase or in the \bba phase) and a public random seed derived from the last published block of the blockchain. 
To prevent Sybil attacks, nodes are weighted by their account balance in the sortition process.
A unit of currency held by a node is considered a sub-node. 
A node with $w$ units of currency is represented by $w$ sub-nodes.
The probability for a sub-node to be chosen is exactly $p = \tau/W$ where $\tau$ is the expected number of nodes chosen for a given phase of a round and $W$ is the total amount of currency units.
The (verifiable random) hash value of the VRF determines how many sub-nodes are selected, and the probability that $k$ of the $w$ sub-nodes follow the binomial distribution:
$B(k;w,p) = (_k^w)p^k(1-p)^{w-k}$ where $\sum_{k=0}^w=0 B(k,w,p) = 1$. Since $B(k_1;n_1,p) + B(k_2;n_2,p) = B(k_1 + k_2;n_1 + n_2,p)$, splitting money across nodes does not provide any gain to an adversary.
If a node is selected between one and $w$ times, it returns a hash value and a proof. 
The proof can then be used when sending messages to other participants to let them validate the correctness of the hash value.





\subsection{Block proposal} 
All nodes execute the cryptographic sortition to determine if they are selected for the block proposal committee of expected size $\tau_{proposer}$. 
Because proposers will all gossip their own proposed block, they also need to determine which block everyone should adopt and 
reduce the unnecessary communication costs as blocks may be as large as a few MBytes. Algorand node sends two messages through the gossip network. 
A first message ($\sim$200 Bytes) includes the node's proof of selection and a priority hash used to identify the priority of a block. 
The latter priority hash is derived from hashing the VRF hash output concatenated with the sub-node index. 
The first message enables most nodes to quickly learn which proposer has the highest priority, and thus discard messages about other proposed blocks that they hear about. 
The second message includes the proposed block that contains transactions.

In each round, block proposers add to their blocks a seed proposal required for the sortition process in the next round. 
The proposal seed is determined as the verifiable random hash value of the last round seed concatenated with current round index $r$, \ie $\langle seed_r,\pi \rangle \leftarrow VRF_{sk}(seed_{r-1} || r)$.
Algorand requires the private key $sk$ of each node to be chosen well in advance of the seed for that round; otherwise, a malicious node could control the seed added to its proposed block.


Before moving on to the next phase of Algorand, nodes should wait long enough to receive block proposals.
Choosing this time interval does not impact Algorand's safety guarantees but is essential for performance. 
A too-short waiting time would result in the absence of block proposals, leading to the initialization of the next phase with an empty block.
Conversely, waiting too long would increase confirmation latency. 
Algorand defines the estimated duration for a node to learn which block has the highest priority once in the block proposal phase. This estimation is composed of the approximated variance in how long it takes for the remaining nodes to finish the last phase of algorand $\lambda_{stepvar}$, and the estimated time for the priority and proof messages of proposers to disseminated $\lambda_{priority}$.
Experiments~\cite{AlgorandGoalgorand2021} show that these parameters are, conservatively, 5 seconds each. 
Once this $\lambda_{stepvar} + \lambda_{priority}$ duration is over, nodes wait to receive the highest priority block within a timeout $\lambda_{block}$ of the order of a minute after which they fall back to an empty block for the initialization of the next phase.




\subsection{Agreement using \ba} 
Block proposal does not guarantee that all nodes receive the same block. 
Algorand relies on a Byzantine Agreement procedure named \ba to reach consensus on a single block.
This procedure is composed of two phases. 
The first one is a reduction, which transforms the problem of agreeing on one block among many to agreeing on either an empty block or a proposed block.
The second phase consists in executing a binary byzantine agreement procedure named \bba.
For efficiency purposes, \ba votes for hashes of blocks instead of entire block contents. 
At the end of \ba, nodes that have only heard about the hash of the already agreed-upon block obtain it from other nodes, and since \ba has terminated, many honest nodes must have received it.

The reduction phase is composed of precisely two steps. The \bba of 2 to 11 steps depending on whether the highest block proposer was an honest node (\ie sent the same block to all users) or colluded with a large fraction of committee participants at every step.
Each node participates in \ba by counting votes, verifying signatures and proofs, and reaching agreement. 
Committee members selected during the cryptographic sortition procedure further contribute by casting votes on block hash values.
An essential aspect in \ba is that a different committee is formed for each step. Indeed, replacing committee participants prevents targeted attacks against honest nodes.
We first describe how nodes count votes before detailing the reduction and \bba procedure. 

\subsubsection{CountVotes Procedure}
In the \bba and reduction phases, nodes are responsible for counting votes.
The \textit{CountVotes} procedure is in charge of ensuring the validity of vote messages and summing up the votes for a given value (\ie a block hash).
Since several sub-nodes of a node can be selected by the sortition process, the \textit{CountVotes} procedure must reflect this by returning the associated number of votes. 
Let $T$ be the \ba election threshold (typically, $T > 2/3$ to satisfy safety) and $\tau$ the expected committee size.
As soon as a value receives more than $T \times \tau$ votes the \textit{CountVotes} procedure returns it, and the node considers that value accepted for the step. 
If no value receives enough votes in a well-defined period of time $\lambda_{STEP}$, the procedure results in a TIMEOUT. 
The \textit{CountVotes} procedure ensures that if a value has been accepted for at least one honest user, then instances of the procedure of other honest users will return either the same value or a TIMEOUT, even under weak synchrony assumption.

\subsubsection{Reduction}
The reduction phase always takes two steps to complete and consists in transforming the problem of reaching consensus on one block hash among many, to reaching consensus on either the hash of a proposed block or the hash of an empty one. 
During the first step, committee members vote for the block hash with the highest priority they have received. 
Suppose a block hash receives at least $T\times\tau$ votes in the first step, then committee members vote for this one in the second step; otherwise, they vote for the hash of a default empty block.
The reduction phase ensures that at most one non-empty block will be selected to be sent to \bba.
Even if a malicious highest-priority proposer leads honest nodes to start the reduction procedure with different blocks, the process will select an empty block.
\subsubsection{\bba}
This second phase consists in reaching consensus on one of the two values passed on to \bba by the reduction phase.
This phase takes a variable (but bounded) number of steps to complete depending on the network synchrony and the honesty of the highest-priority proposer. 
Nodes continue to validate messages and count votes as in the reduction phase, and the committee members of each step cast votes on block hashes.
The voting scheme remains the same as in the previous phase.





In each step of \bba, when a node receives more than $T \times \tau$ votes for some value, it votes for that same value in the next step.
In the common case, when the network is strongly synchronous and the block proposer is honest, \bba starts with the same block hash for most nodes, and will reach consensus in the second step, since most committee members vote for the same block hash.
However, if no value receives enough votes, \bba chooses the next vote in a way that ensures consensus in a strongly synchronous network.

If the network is not strongly synchronous (\eg there is a partition), \bba may return consensus on two different blocks. 
\ba addresses this problem by introducing the notion of \textit{final} and \textit{tentative} consensus. 
Final consensus means that \ba will not reach consensus on any other block for that round. 
In that event, transactions contained in the agreed-upon block are considered as confirmed. 
On the other hand, tentative consensus informs that \ba could not guarantee safety, that other nodes may have reached tentative consensus on a different block, and that nodes must rely on the next rounds for this safety on the current block(s).
Transactions contained in a tentative consensus block are only confirmed when a final consensus is reached on a successor block (or after the execution of a recovery described in the original Algorand paper \cite{AlgorandGoalgorand2021}).

\ba produces tentative consensus in two cases. 
First, if the network is strongly synchronous, an adversary may, with small probability, cause \ba to reach tentative consensus on a block. 
In this case, \ba will not reach consensus on two different blocks but is simply unable to confirm that the network was strongly synchronous. 
Algorand will eventually (in a few rounds) reach final consensus on a successor block, with overwhelming probability, and thus confirm these earlier transactions.

The second case is that the network was only weakly synchronous (\ie it was entirely controlled by the adversary, with an upper bound on how long the adversary can keep control). 
In this case, \ba can reach tentative consensus on two different blocks, forming multiple forks. 
This can prevent \ba from reaching consensus again because the users are split into different groups that disagree about previous blocks. 
To recover liveness, Algorand periodically invokes \ba to reach consensus on which fork should be used going forward. 
Once the network regains strong synchrony, Algorand chooses one of the forks, and then reach final consensus on a subsequent block on that fork.


\subsubsection{Committee size}

The expected size $\tau$ of committees in \ba and the fraction $T > 2/3$ of votes needed to reach consensus are crucial parameters for safety, liveness and performance.
From the proportion $h>2/3$ of money held by honest nodes, committees must be composed of sufficiently honest nodes for safety.
Due to the probabilistic property of cryptographic sortition, there is a small chance for committees to be composed of a little less than $T\times \tau$ honest nodes, hence we would like to have $T$ as high as possible.
However, large committees translate into higher bandwidth cost and latency.
Reducing the committee size could benefit performance and liveness, but requires to increase $\tau$ to ensure that an adversary does not obtain enough votes by chance. 
\ba's goal is to make this probability negligible. 
Two different parameter sets are considered: $T_{final}$ and $\tau_{final}$ for the final step of \ba, which ensures an overwhelming probability of safety regardless of strong synchrony, and $T_{step}$ and $\tau_{step}$ for all other steps, which achieve a reasonable trade-off between liveness, safety, and performance.
Algorand's proofs of safety, liveness and performance~\cite{giladAlgorandScalingByzantine2017a} describe that at $h = 80\%$, $\tau_{step}$ = 2000 can ensure that safety holds with probability $1 - 5 \times 10^{-9}$ (using $T_{step} = 0.685$).
Under weak synchrony, $\tau_{final} = 10 000$ suffices with $T_{final} = 0.74$.





%% file: sections/overview.tex
\section{\protoname}\label{sec:approach}


\protoname multiplexes several instances of Algorand and inherits its safety and liveness properties.
Building on top of Algorand, \protoname allows multiple nodes to independently propose concurrent blocks containing disjoint sets of transactions and grows the chain by appending a subset of the proposed blocks in every round. 

\begin{figure}[ht]
  \centering
  \includegraphics[width=\columnwidth]{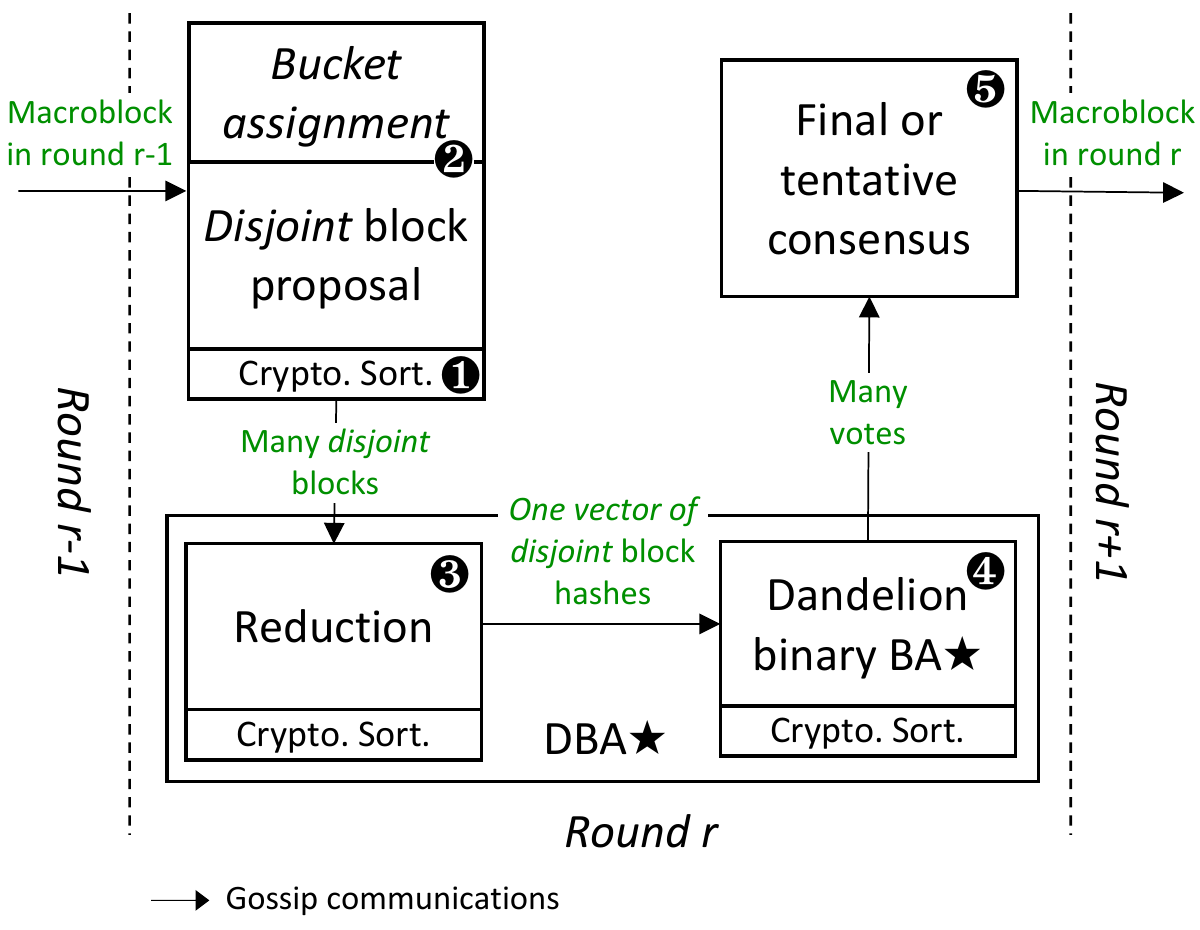}
  \caption{Structure of a round in \protoname}
  \label{fig:dand_round}
\end{figure}

Figure~\ref{fig:dand_round} depicts the global course of a round.
First, each node executes the cryptographic sortition to determine whether it is elected to contribute to the block proposal phase (step \dingnum{1}) as well as the other phases of \dba.
Then, it locally builds a block in a way that ensures that at least one other proposed block will contain non-colluding transactions (step \dingnum{2}).
Then, it locally builds a block in a way that ensures proposers can create blocks composed of disjoint sets of transactions(step \dingnum{2}).
Nodes gossip blocks through the network before initiating the reduction phase that produces either a vector of disjoint block hashes or the hash of an empty block in precisely two steps (step \dingnum{3}).
Finally, the Dandelion binary byzantine agreement \dbba operates between 2 to 11 steps to decide on a final or tentative consensus as in Algorand (steps \dingnum{4} and \dingnum{5}).
Nodes can then locally gather the blocks \protoname has agreed on and deterministically build the composition of these blocks that we call \textit{macroblock}.

\subsection{Structure of a macroblock}
A macroblock is a logical composition of up to $Cl$ blocks of size $bs$, with $Cl$ a bootstrap parameter that defines the concurrency level of \protoname. 
Consequently, \protoname grows a chain of macroblocks of possibly different sizes.
The \dba agreement protocol produces a vector of block hashes that compose the macroblock for the current round.
Based on this decision, each node can locally build the agreed-upon macroblock by concatenating the associated blocks and the vector of block hashes. 
In most blockchains, blocks include the hash of their predecessor, creating a chain of blocks. 
\protoname's chain is slightly different because each block needs to refer to all the blocks composing the preceding macroblock. 
For this reason, each block contains the hash of the previous macroblock.

\begin{figure}[ht]
  \includegraphics[width=\columnwidth]{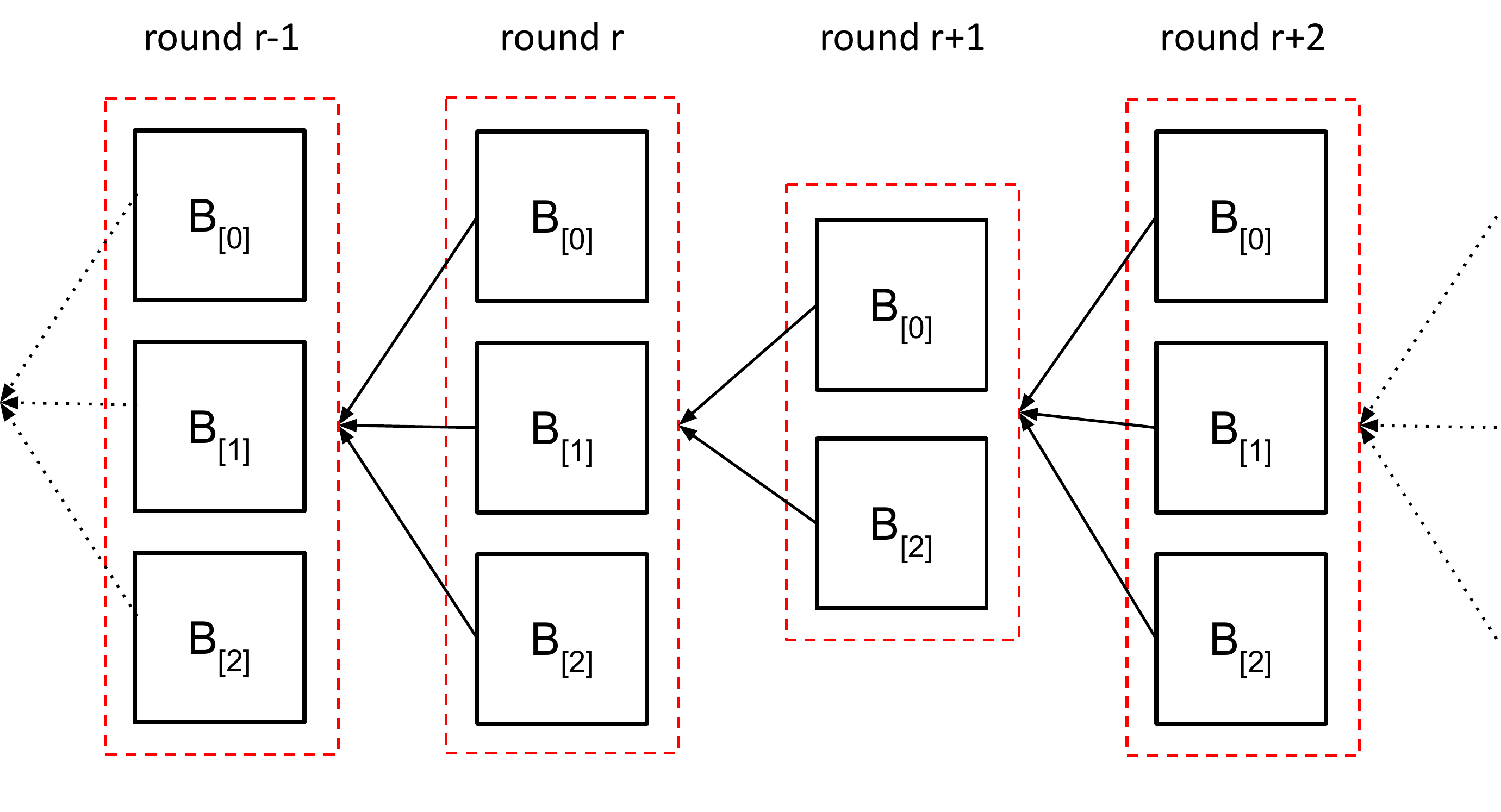}
  \caption{Structure of Dandelion's blockchain}
  \label{fig:dandelion_blockchain_structure}
\end{figure}

Figure~\ref{fig:dandelion_blockchain_structure} depicts an example of a chain of macroblocks with a concurrency level $Cl$ of 3.
Dashed lines represent macroblocks, and solid squares show the composing blocks.
The macroblock in round $r$ is composed of 3 blocks, whereas the next macroblock contains 2 block only. 
Although very unlikely a macrocklock could include no bock at all, in that case the macroblock is composed of a default empty block that contains a null payload of transactions.

\subsection{Transaction duplication and duplication attacks}

Appending multiple blocks in a given round poses the challenges of transaction duplication. 
Namely, in a simple approach, proposers could create blocks with any available transaction they receive.
As a result some blocks may contain transactions that have already be appended to other blocks, which would reduce the throughput gain envisioned in our approach and increase the complexity of the transaction execution phase.
Elected byzantine nodes could also wait to learn about blocks proposed by honest nodes and form a block with the same transactions in it.
Even worse, Byzantine nodes could produce blocks composed of identical fake transactions, which would drastically slash the performance of \protoname and annihilate the interest of appending several blocks in a round.
To cope with these problems, and inspired by the technique employed in Mir-BFT~\cite{stathakopoulouMirBFTHighThroughputBFT2019a}, \protoname relies on a transaction hash space partitioned into buckets.
Specifically, buckets are assigned in a verifiable random manner to block proposers, hence preventing request duplication.

\subsection{Transaction partitioning and bucket assignment}\label{sec:bucket-assignment}

\protoname partitions the transaction hash space into $Cl$ non-intersecting buckets of approximately equal size. 
Each block proposer is assigned a bucket and can include in its block proposal the transactions which hashes fall into its bucket only.

\begin{figure}[ht]
  \centering
  \includegraphics[width=.55\columnwidth]{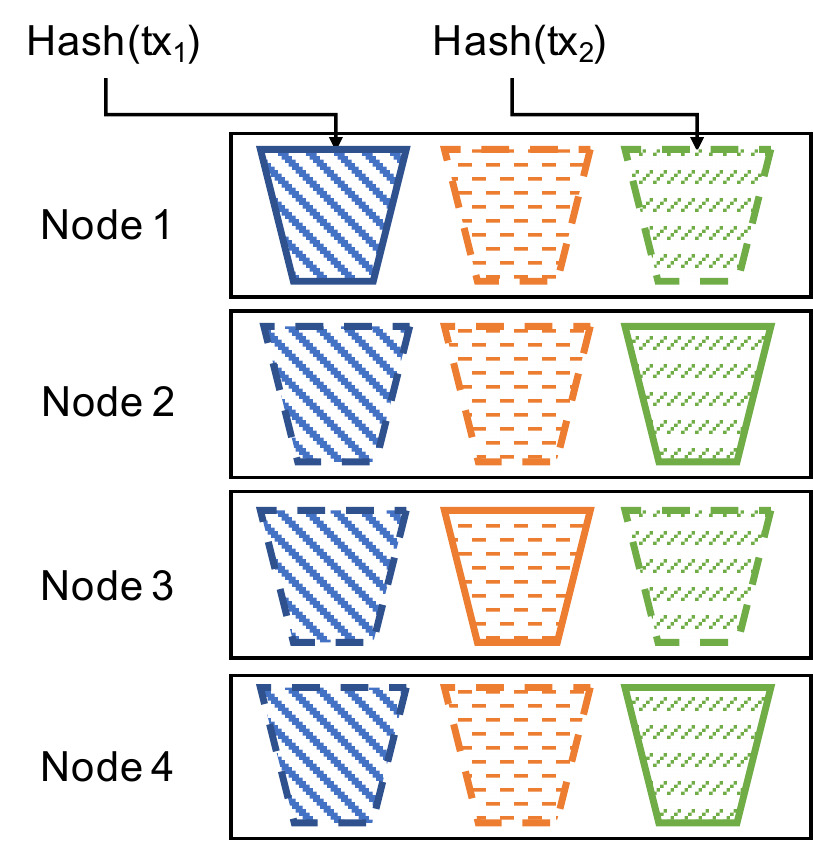}
  \caption{Mapping transactions to bucket with $Cl=3$. Solid lines represent the assigned bucket for a given node. Hash(tx$_1$) is mapped to the first bucket, assigned to node $i$. Hash(tx$_2$) is mapped to the third bucket, assigned in node 2 and 4}
\label{fig:bucket}
\end{figure}

Figure~\ref{fig:bucket} illustrates the mapping of transactions to assigned buckets with $Cl=3$.
The bucket assignment is deterministic and publicly verifiable so that any node in the system can validate a block issued by a given elected node.
In \protoname, the elected nodes determine their assigned bucket by using the VRF's hash output returned during the cryptographic sortition process.
The assigned bucket index is computed from the VRF's hash modulo the concurrency level $Cl$.
Because \protoname decides on a vector of blocks and not one block only, the number of block proposers $\tau_{proposer}$ should be sufficiently large to avoid unassigned buckets.
A too-low $\tau_{proposer}$ value could lead to buckets being frequently unassigned which would not only hinder the envisioned throughput gain, but would also lead to transactions with specific hash being unfairly ignored for some period.
To devise an appropriate value for $\tau_{proposer}$, we rely on the uniform distribution of bucket assignment deriving from the use of hash functions in the cryptographic sortition.
Namely, each node has equal chances of being assigned a bucket over another one; a bucket are assigned to at least one proposer with probability $1 - (1-\frac{1}{Cl})^{\tau_{proposer}}$; and all buckets are assigned at least one proposer with probability $\sum_{i=0}^{Cl} (-1)^{Cl-i}\binom{Cl}{i} \big( \frac{i}{Cl} \big)^{\tau_{proposer}}$.


  
\subsection{Disjoint block proposal and seed generation}  

The block proposal phase of \protoname is very similar to that of Algorand, except that nodes can be assigned the same transaction bucket as other nodes.
Similarly to Algorand, we employ two kinds of messages for nodes to decide which block everyone should adopt for each bucket and which block can be safely discarded. 
The first message is composed of the proof of selection and a priority hash value used to compare blocks with transactions issued from the same bucket index. 
The priority hash is derived from hashing the VRF hash output concatenated with the sub-node index and the bucket index.
The second message is the block itself.

The seed generation mechanism is also different.
Indeed, in Algorand, block proposers add to their blocks a seed proposal required for the sortition process in the next round. 
This proposed seed is determined as the verifiable random hash value of the last round seed concatenated with current round index $r$, \ie $\langle seed_r,\pi \rangle \leftarrow VRF_{sk}(seed_{r-1} || r)$.
In \protoname, each elected node proposes a partial seed by executing the VRF function on the concatenation of the seeds from all blocks in the previous macroblocks. 
That is $\langle seed_r,\pi \rangle \leftarrow VFR_{sk}(seed_{r-1,1}||...||seed_{r-1,Cl}||r)$. 
Doing so, the next round seed is only revealed when \dba decides on the composition of the next macroblock.
Then, the actual seed used to run crytographic sortition is the hash the seeds from all the blocks composing the last macroblock.
\subsection{\dba}

Before moving on the \dba, each node waits $\lambda_{priority} + \lambda_{stepvar}$ to receive enough block propositions and priority messages for each bucket.
At the end of this time interval, each node creates a candidate vector $V_h=\{hash_i\}_{i \in [1 \isep Cl]}$ of $Cl$ hashes of received blocks. 
$V_h$ identifies a candidate macroblock in the \dba procedure.
In each phase of \dba, nodes vote for a vector of block hashes and nodes continue voting until one of the proposed vectors reaches a threshold vote count.
\protoname relies on the same threshold count $T.\tau$. 


%% file: sections/implementation.tex
\section{Implementation}\label{sec:experimentation}


We implemented a prototype of \protoname in Golang, consisting of approximately 3,000 lines of code. 
We rely on the official Algorand VRF implementation~\cite{AlgorandGoalgorand2021} and on the SHA-256 hash function.
In our implementation, each node selects and connects to 4 random peers, accepts incoming connections from other peers, and gossips messages to all of them.
Therefore, each node communicates with 8 peers on average. 
To simplify the experimental process, each node in the system knows the IP addresses and port numbers of all other peers.
In a real-world deployment, a Sybil-resilient peer discovery service would be necessary. 
However, this problem is orthogonal to our approach, and existing solutions already exist~\cite{piletFoilingSybilsHAPS2020a}.

\begin{table}[ht]
    \centering
    \caption{Protocol parameters of Algorand and \protoname}
    \begin{tabular}{l l l} 
        \noalign{\hrule height .7pt}
        Parameter & Meaning & Value  \\  [0.5ex] 
        \noalign{\hrule height .6pt}
        $h$ & assumed fraction of honest weighted users & 80\% \\  
        $\tau_{proposer}$ & expected \# of block proposers & 100 \\ 
        $\tau_{step}$ & expected \# of committee members & 2,000  \\
        $\tau_{final}$ & expected \# of final committee members & 10,000  \\
        $T_{step}$ & threshold of $T_{STEP}$ for BA$\star$ & 68.5\%  \\
        $T_{final}$ &  threshold of $\tau_{FINAL}$ for BA$\star$ & 74\%  \\
        $\lambda_{priority}$ & time to gossip sortition proofs & 5 seconds  \\
        $\lambda_{block}$ & timeout for receiving a block & 120 seconds \\
        $\lambda_{step}$ & timeout for BA$\star$ step & 20  seconds \\
        $\lambda_{stepvar}$ & estimate of BA$\star$ completion time variance & 5 seconds  \\
        \noalign{\hrule height .6pt}
    \end{tabular}
    \label{table:protocol_parameters}
\end{table}

\begin{figure}[ht]
    \center
    \includegraphics[width=\columnwidth]{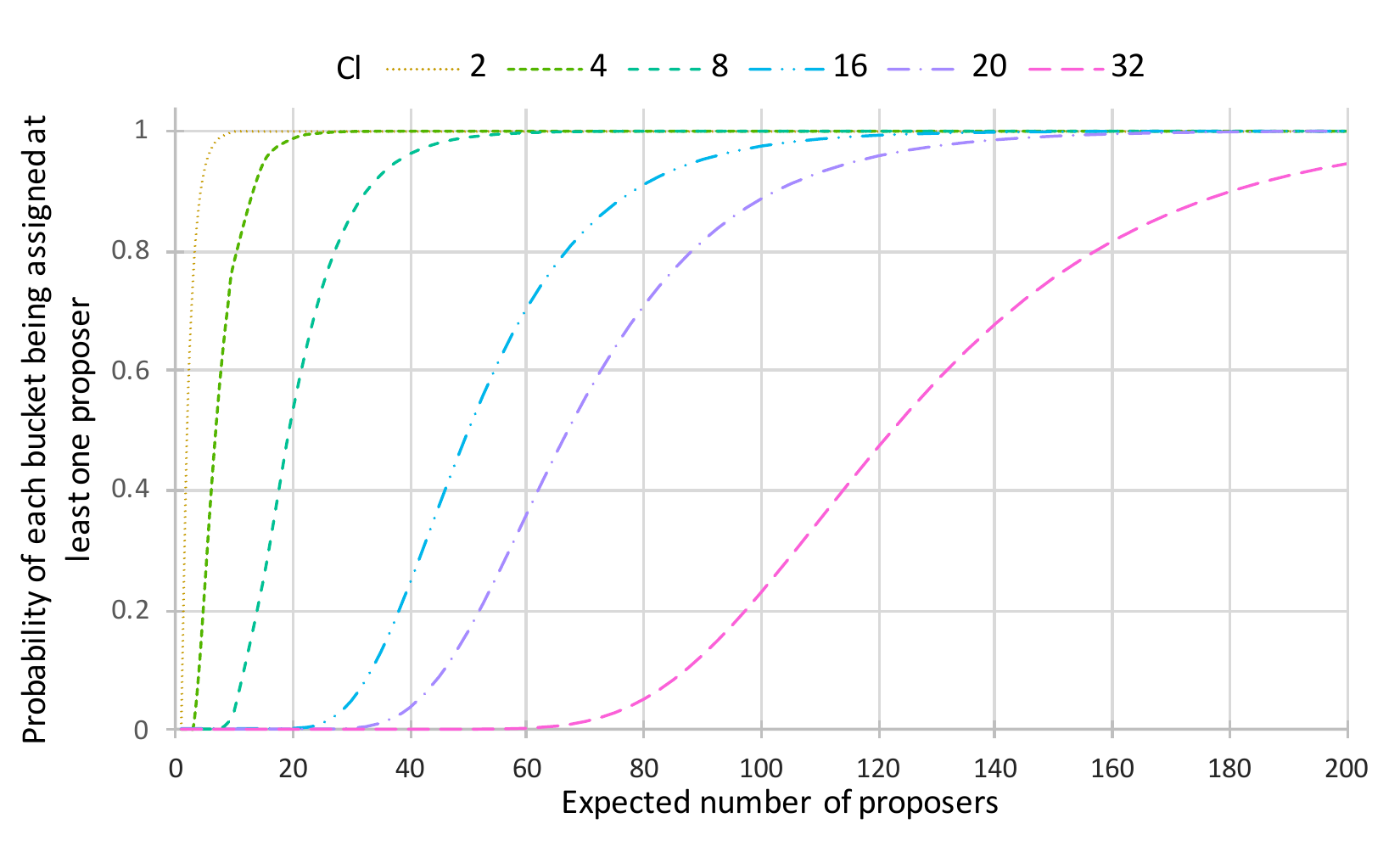}
    \caption{Propability that each bucket will be assigned at least one proposer}
    \label{fig:taux-proposer}
\end{figure}

Table~\ref{table:protocol_parameters} shows the parameters used in our prototype of \protoname.
Although most of them are identical to those of the original Algorand experiments~\cite{AlgorandGoalgorand2021}, we modify two parameters to explore performance and scalability in depth: the expected number of block proposers, and the timeout for receiving a block. 
We set the expected number of block proposers $\tau_{proposer}$ to 100, instead of 26. 
To find this value, we computed the probability that each bucket is assigned at least one proposer (see Section~\ref{sec:bucket-assignment}) when varying the expected number of proposers and the concurrency level. 
As shown in Figure~\ref{fig:taux-proposer}, using $\tau_{proposer}=100$ ensures that this probability is very high (greater than $90\%$) with almost all configurations, except for $Cl=32$.

The second modified parameter is $\lambda_{block}$,  which controls how long a node waits to receive the block with the highest priority it has heard about. 
Beyond that time limit, the node initiates the \dba or \ba agreement with the hash of an empty block.
Since we use large block sizes to evaluate \protoname and Algorand, the block propagation time can be  significantly increased.
Therefore, we set $\lambda_{block}$ to 120 seconds instead of 60, to avoid prematurely compromising the smooth execution of the protocol.

%% file: sections/evaluation.tex
\section{Evaluation}\label{sec:evaluation}

We deploy our prototype of \protoname on Grid'5000, a large-scale and flexible testbed for experiment-driven research.
We use up to 100 large physical machines, each of which has 18 cores, 96GB of memory, and a 25 Gbps intra network connectivity link.
To measure the performance with a large number of nodes, we run 100 \protoname nodes on each physical machine. 
To simulate commodity network links, we cap the bandwidth for each \protoname process to 20~Mbps using traffic control rules.
Similarly to Algorand, we use inter-city latency and jitter measurements~\cite{pingTimes} to model network latency. 
We assign each machine to a city and add a one-way latency of 50 milliseconds to each machine; latency within the same city is modeled as negligible.

We assigned an equal share of money to each node to maximize the number of messages during each round.
To start the experiment, each node initiates the \protoname protocol for 17 rounds. 
We then collect the results and keep only data between rounds 5 and 15.
In the remainder, graphs plot the time it takes for \protoname to complete an entire round, and include the minimum, median, maximum, 25th, and 75th percentile times across all nodes and all rounds.

\subsection{System performance}

We deploy 1,000 nodes on ten machines (100 nodes per machine) and vary the macroblock size from 1~MB to 24~MB and the concurrency level from 1 to 32. 
The experiments with the concurrency level 1 ($Cl=1$) exhibit the behavior of Algorand.

\begin{figure}[ht]
    \centering
    \includegraphics[width=\columnwidth]{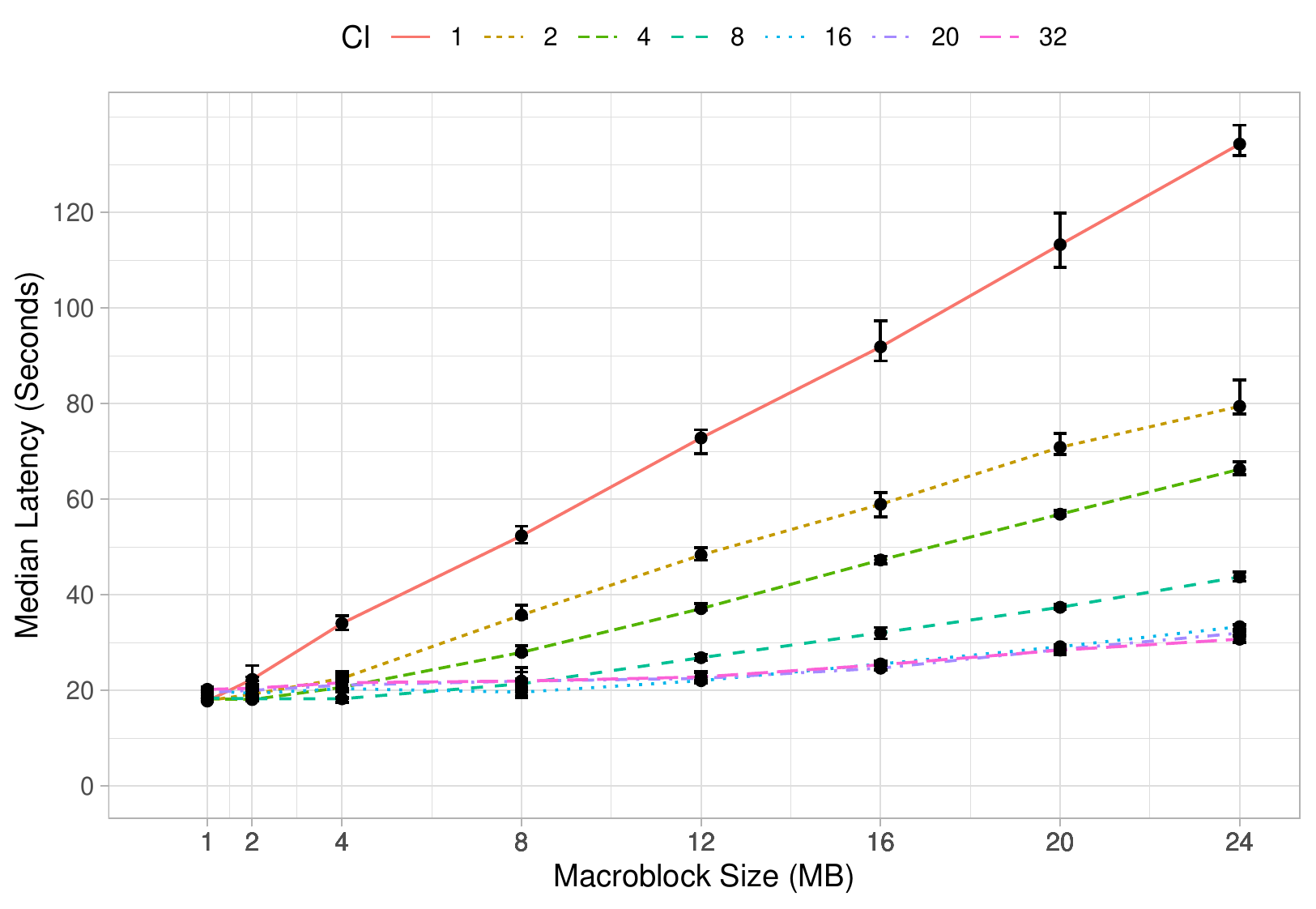}
    \caption{Round Latency}\label{fig:round-latency}
\end{figure}

We consider protocol round latency, which indicates how long it takes for a block to be appended to the chain.
We evaluate the duration between the proposal of a block by a node and the point where all nodes observe this block in the blockchain.
As shown in Figure~\ref{fig:round-latency}, Algorand's round latency rapidly grows with the size of the blocks appended. 
It largely dominates any configuration of \protoname with macroblock size larger or equal to 2~MB.
For example, Algorand needs 110 seconds to append a 20~MB block.
In contrast, under $Cl$ values between 16 and 32, \protoname shows a much lower limit on the round latency that it achieves, about 30 seconds for macroblocks of 20~MB.

\begin{figure}[ht]
    \centering
    \includegraphics[width=\columnwidth]{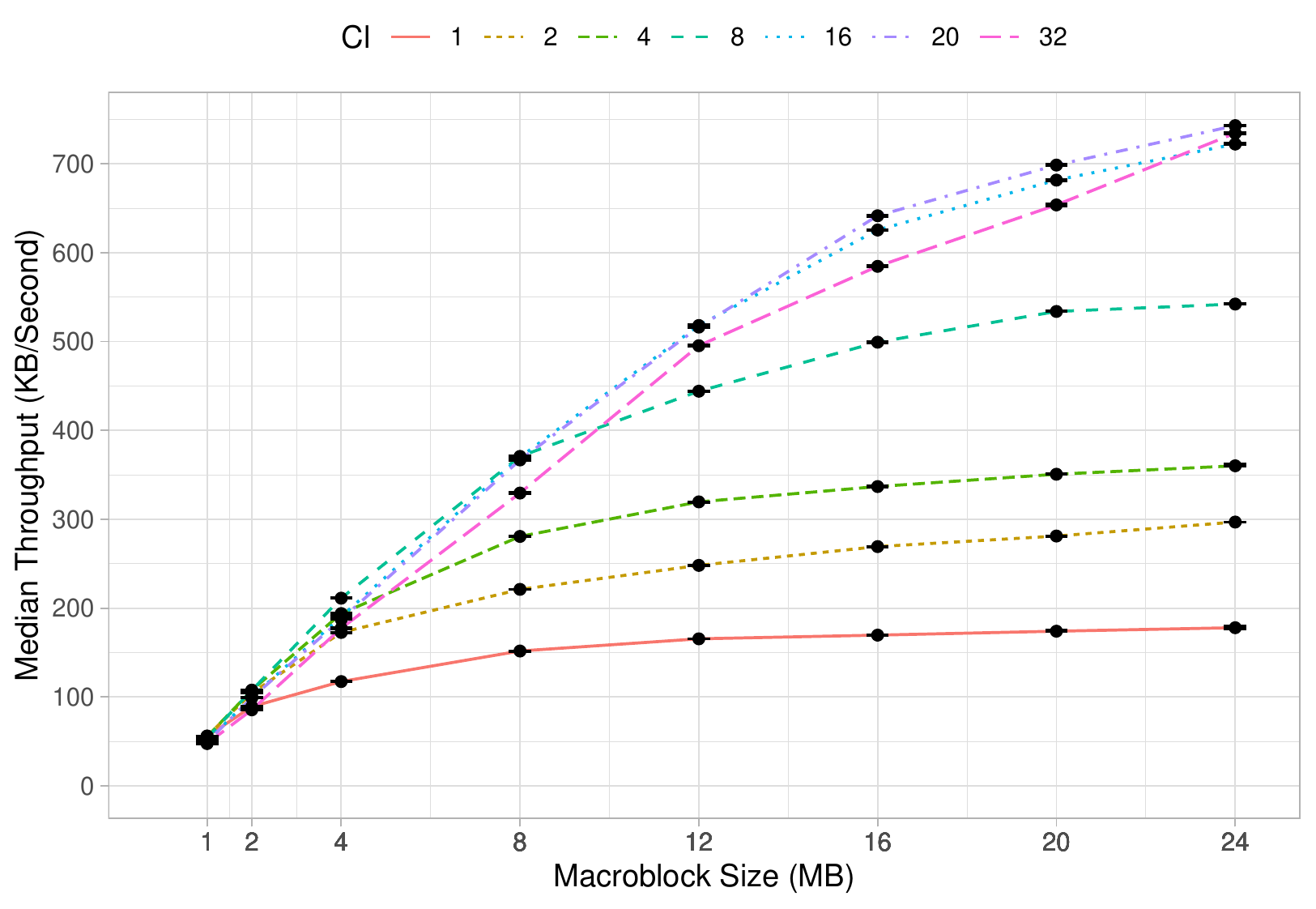}
    \caption{Effective throughput}\label{fig:effective-throughput}
\end{figure}

We evaluate the effective throughput of \protoname.
Since some macroblocks may contain fewer blocks than expected, we cannot derive throughput directly from the round latency and the macroblock size. 
We define the effective throughput as the actual amount of data per second appended to the blockchain. 
As shown in Figure~\ref{fig:effective-throughput}, \protoname outperforms Algorand in all configurations and can reach up to 740KB/s, a 4 times improvement.

Our experiments highlight the limits of multiplexing Algorand instances.
Indeed, we reach the maximum throughput with $Cl=16$ and increasing the concurrency level does not improve performance further. 
In addition, the configuration with $Cl=20$ provides a similar if not slightly better throughput than the one with $Cl=32$.

There are two possible reasons for these limitations.
First, the expected number of block proposers ($\tau_{proposer}$) in the 32-$Cl$ configuration is too low, resulting in more unassigned blocks than the 20-$Cl$ configuration and preventing throughput improvement.
Second, the network usage by the gossip dissemination of blocks has reached its limits; some blocks cannot be transmitted fast enough, leading many nodes to initiate \dba with empty block hash entries in the block hash vector. 
As a result, the effective throughput can no longer be substantially improved.

We perform the following experiments to investigate the root cause of these observed limitations. 
We run \protoname with $Cl=32$ with macrobocks of 24~MB, 28~MB, and 32~MB.
For each configuration, we vary the  expected number of block proposers from 100 to 200. 
As shown in table~\ref{table:tauprop}, the effective throughput for a given macroblock size varies insignificantly with the expected number of proposers.
As a conclusion, \protoname is already using the maximum of the gossip network and cannot reach higher throughput.

\begin{table}[ht]
    \centering
    \caption{Impact of $\tau_{proposer}$ on the effective throughput}
    \begin{tabular}{c c c c} 
        \midrule
        Macroblock  & throughput for & throughput for  & throughput for  \\
        size &  $\tau_{prop.} = 100$ & $\tau_{prop.} = 150$ & $\tau_{prop.} = 200$\\
        \midrule
        24~MB & 732 KB/s & 745 KB/s& 738 KB/s\\ 
        28~MB & 781 KB/s & 789 KB/s& 785 KB/s\\ 
        32~MB & 809 KB/s & 804 KB/s& 802 KB/s\\ 
        \midrule
    \end{tabular}
    \label{table:tauprop}
\end{table} 

\subsection{Scalability}
\begin{figure}[t]
    \includegraphics[width=\columnwidth]{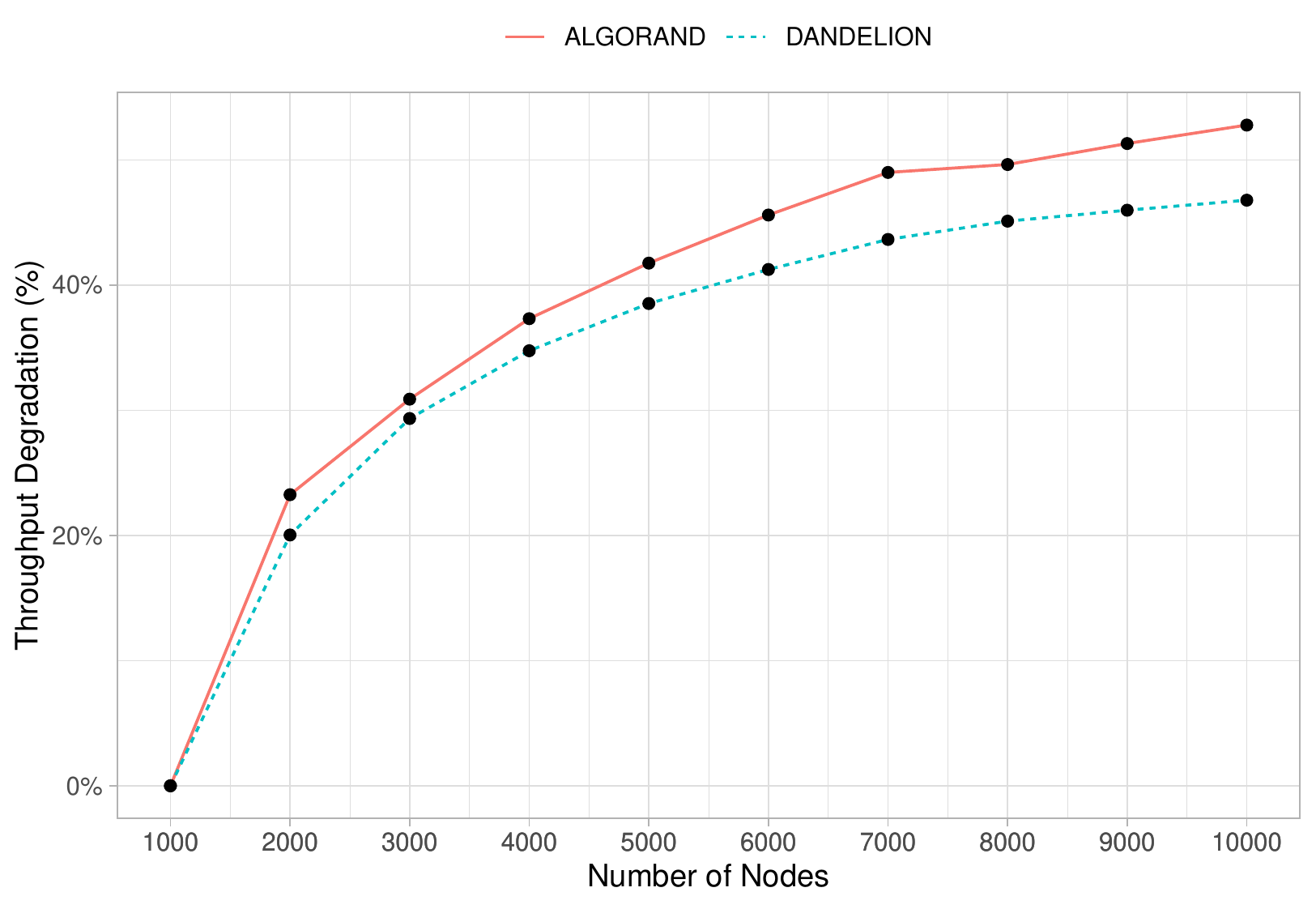}
    \caption{Effective throughput degradation}
    \label{res:scale_thr}
    
    \includegraphics[width=\columnwidth]{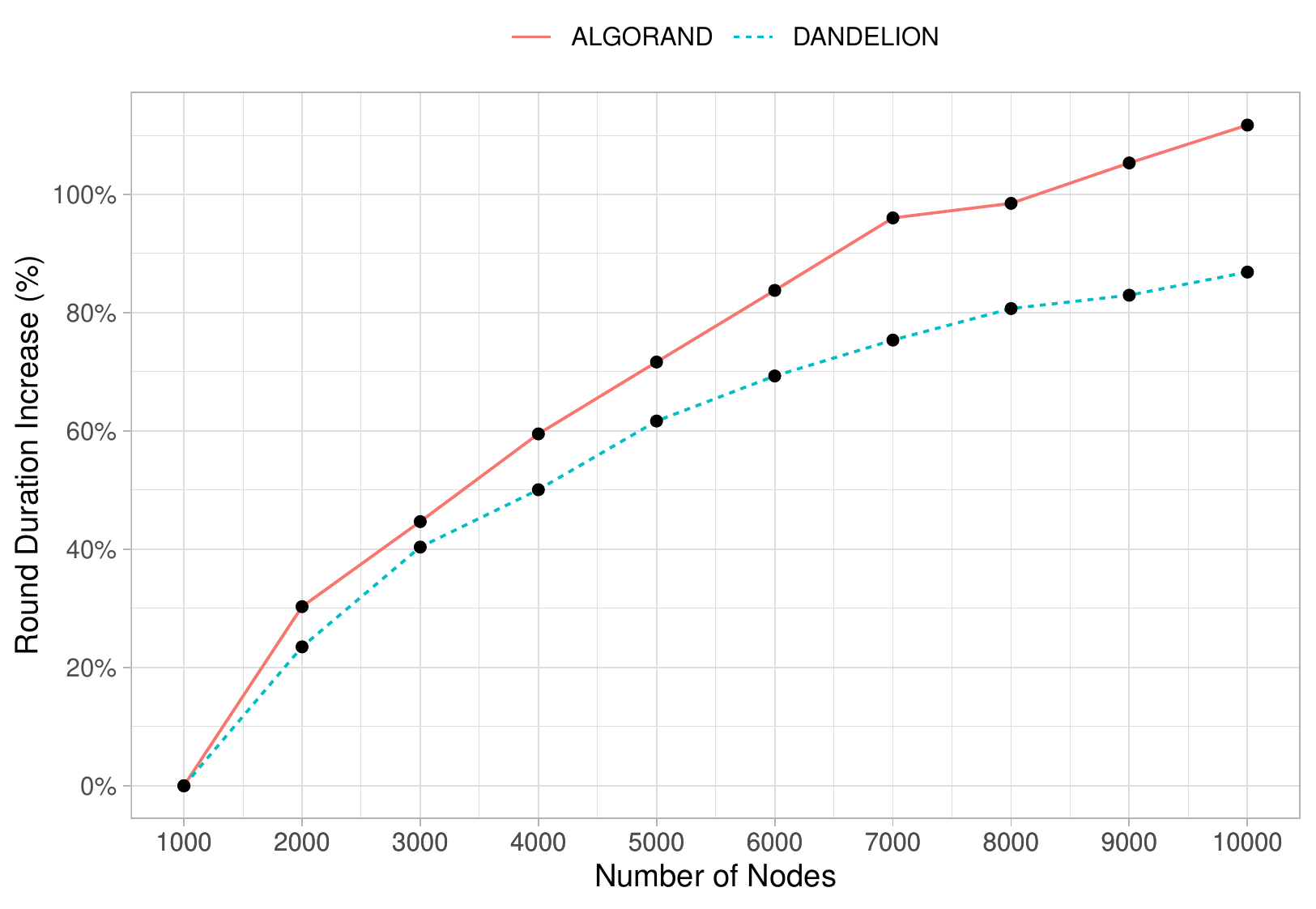}
    \caption{Round duration increase}
    \label{res:scale_lat}
\end{figure}

To evaluate how \protoname and Algorand scale, we vary the number of machines  in the testbed from 10 to 100 with 100 nodes per machine, allowing us to emulate up to 10,000 nodes.
Figure~\ref{res:scale_thr} illustrates the throughput degradation when the number of nodes increase, compared to a baseline with 1,000 nodes.
The latency increase is shown in Figure~\ref{res:scale_lat} and we can observe that \protoname scales as well as if not better than Algorand.

\subsection{Application-specific performance}

In our experiments, we focus on system performance metrics only by abstracting from the transaction data structure itself.
We do not consider optimizations that aim at improving the number of transactions per second (tps) from the application point of view.
Existing approaches, such as the one currently followed by Algorand~\cite{AlgorandPerformance20201}, consist in building blocks of 32-byte transaction hash instead of the average 500-byte transaction size, and even lowering the transaction footprint inside block to a few bits only. 
Doing so, the new performance report of Algorand claims to achieve 46,000 tps in 2021.
Our approach is complementary to these extensions and would result in even better performance with \protoname, paving the way for reaching 200,000 tps.

%% file: sections/sota.tex
\section{Related work}
\label{sec:rw}

This section reviews the different PoS blockchain approaches in the literature and how they address the Sybil resilience.
The idea of PoS blockchains comes from the Bitcoin community as an energy-efficient alternative to PoW mining. 
Essentially, the stake refers to the number of coins belonging to a participant that can be invested in the blockchain consensus protocol. 
PoS leverages token ownership to address Sybil attacks. 
Compared to a PoW protocol where a node's probability of proposing a block is proportional to its computation power, in a PoS protocol, its chances are proportional to its stake in the system. 
From the node's perspective, PoS shifts its involvement from outside the system \ie computation power and electricity, to inside the system \ie loss of money.
Following the taxonomy acknowledged in the literature \cite{xiaoSurveyDistributedConsensus22,huangSurveyStateoftheArtBlockchains2021,banoSoKConsensusAge2019}, we review four classes of PoS protocols: chain-based PoS, committee-based PoS, Byzantine-fault-tolerant-based (or BFT) PoS, and delegated PoS (DPoS). 

\subsection{Chained based PoS}
The Chain-based PoS approach is the first PoS scheme proposed as an alternative block generation mechanism to PoW. 
Some of the first chain-based PoS blockchain systems include Peercoin\cite{kingPPCoinPeertoPeerCryptoCurrency} and Nxt~\cite{NxtWhitepaperIntroduction}.
The general block generation procedure in a chain-based PoS can be summarized as follows.
Nodes join the network by connecting to known peers and seal some coins in the stake pool.
Each node gathers available transactions and prepares a block header, including the transaction Merkle-tree root.
Then, each node attempts to find a hash value starting with a target number of zero bits derived from hashing the block header concatenated with the clock time. 
The target difficulty is weighted by the number of sealed coins.
Since the hashing puzzle difficulty decreases with the node's stake value, the expected number of hashing attempts for a node to solve the puzzle can be significantly reduced if its stake value is high. 
Therefore, PoS avoids the hashing competition of PoW-based protocols, achieving a significant reduction of energy consumption.

Although confirming blocks in the order of hours, chain-based PoS can tolerate up to 50\% of all stakes being controlled by the adversary.
If colluding attackers control more than 50\% of the stake, they can grow their malicious chain faster than the others and double-spend.
However, PoS attackers have lower incentives to do so because of the stake loss risk. 
Nodes can detect users violating the consensus and slash their stake or prevent them from participating in the future staking process.

\subsection{Committee-based PoS}

Committee-based PoS blockchains rely on multi-party computation (MPC) schemes to elect a committee of nodes that generates blocks in a predetermined order.
A secure MPC scheme is often used to derive such a committee in a distributed network.
The MPC essentially outputs a pseudo-random sequence of leaders, subsequently populating the block-proposing committee.
This leader sequence should be the same for all nodes, and those with higher stake values may take up more slots in the sequence. 
Well-known committee-based PoS schemes include Ouroboros~\cite{kiayiasOuroborosProvablySecure2017}, Ouroboros Praos~\cite{davidOuroborosPraosAdaptivelySecure2018}, Bentov’s chain ~\cite{bentoviddoProofActivityExtending2014}, Snow White~\cite{daianSnowWhiteRobustly2019}, Dfinity \cite{hanke2018dfinity}, and Algorand~\cite{AlgorandGoalgorand2021}. These protocols validates block of transactions in the order of minutes.

Ouroboros is a synchronous provably secure PoS algorithm used in the Cardano platform~\cite{cardanoCardanoProject} and proven to be secure against PoS attack vectors.
A random subset of all stakeholders run a multiparty coin-tossing protocol to agree on a random seed. 
The nodes then feed this seed to a verifiable random function that elects a leader from among the participants in proportion to their stake. 
The next set of participants is then elected from the same random seed.

In Ouroboros Praos and Snow-White participants independently determine if they have been elected. 
Snow-White adopts a similar approach to Bitcoin for the leader election. 
Nodes try to find the image of a block that produces a hash below some target. 
However, participants can only compute one hash unit of time and the target takes into account each participant’s amount of stake.

Dfinit is another committee-based PoS based protocol proposed to solve performance issues of existing PoS blockchain systems. Similar to Algorand and \protoname, Dfinity relies on a random beacon and verifiable random functions to select leaders from registered consensus participants. 
The randomness beacon of Dfinity is implemented on top of a threshold signature scheme. 
This construction prevents attackers to learn the identities of leaders in advance. 
Dfinity deploys a block notarization technique on top of the blockchain layer so that a single block is appended to the blockchain on each round.

Despite the use of an ordered block proposing scheme, committee-based PoS blockchains still rely on the longest chain rule for probabilistic finality with an adversary holding less than 50\% of the stakes.
The round-based committee election process ~\cite{kiayiasOuroborosProvablySecure2017} also faces scalability problems.
Large committee sizes may lead to never-ending consensus instances due to the excessive communication costs. 
The duration of rounds can be extended to ensure that all messages are delivered before the nodes proceed to the next round. 
This, however, leads to longer transaction confirmation latency and lower throughput.

\subsection{BFT-based PoS}
BFT-based PoS blockchains combine staking with BFT consensus algorithms to provide fast block confirmation and deterministic finality instead of the probabilistic guarantees of chain-based and committee-based PoS. 
Proposing blocks are performed via any PoS mechanism (round-robin, committee-based, \etc) as long as the blocks are injected at a high enough rate into the BFT consensus protocol.
Aside from the general procedure, a checkpointing mechanism can be used to seal the finality of the blockchain. 
As a result, the longest-chain rule is replaced by the most recent stable checkpoint rule for determining the stable main chain. 
Popular BFT-based PoS blockchain protocols include Tendermint~\cite{kwonTendermintConsensusMininga}, Casper FFG~\cite{buterinCasperFriendlyFinality2017} and GRANDPA~\cite{stewartGRANDPAByzantineFinality2020a}. 

BFT-based PoS’s consensus fault tolerance varies among the implementations mentioned above. 
In Tendermint, although block proposers are determined based on PoS, all nodes have equal weight in the consensus process. 
Therefore Tendermint tolerates up to 1/3 of Byzantine nodes.
In comparison, Algorand, Casper FFG, and GRANDPA tolerate up to 1/3 of maliciously possessed coins. 

\subsection{Delegated PoS}
DPoS employs a voting mechanism that elects a group of so-called delegates missioned to validate transactions and execute the consensus protocol on behalf of the voters. 
DPoS can be thought of as a democratic committee-based PoS as the committee is chosen via public stake delegation. 
It is currently used by EOSIO~\cite{esio.ioEOSIOTechnicalWhitepaper2018}, and Cosmos~\cite{cosmos}. 
DPoS was designed to control the size of the consensus group so that the messaging overhead of the consensus protocol remains manageable. 
A token holder entrusts the delegate with its own stake by casting a vote to a delegate via a blockchain transaction. 
As a result, the delegate represents the stake voting power of its voters and acts as their proxy in the consensus process.
Assuming that the delegates use BFT protocols for block finalization, DPoS can tolerate up to 1/3 of delegates being malicious.

%% file: sections/conclusion.tex
\section{Conclusion}\label{sec:conclusion}

Proof-of-Stake (PoS) based blockchain protocols are gaining popularity as they provide scalability, denial of service resistance, and Sybil resistance while being more energy-efficient than their Proof-of-Work (PoW) based counterparts. 

In this paper, we presented \protoname, a solution that improves Algorand, one of the most effective PoS blockchains. 
Our solution relies on two key design principles (1) allowing multiple nodes to propose verifiably disjoint blocks (2) leveraging a subset of the proposed blocks to build macroblocks. 

We implemented \protoname in Go and measured its performance compared to Algorand on a cluster involving up to ten thousand nodes deployed one up to one hundred machines. 
Results showed that \protoname improves the throughput and latency of Algorand four-fold without changing its original assumptions. 
In our future work, we plan to further abstract the principle of multiplexing consensus protocols from the blockchain protocol itself and apply this methodology to a wide scope of PoW and PoS blockchains. 